# Research trends in combinatorial optimisation

Jann Michael Weinand[1], Kenneth Sörensen[2], Pablo San Segundo[3], Max Kleinebrahm[1], Russell McKenna[4]

[1] Chair of Energy Economics, Institute for Industrial Production, Karlsruhe Institute of Technology, Germany

[2] Department of Engineering Management, University of Antwerp, 2000 Antwerp, Belgium

[3] Universidad Politécnica de Madrid, Centre for Automation and Robotics, Madrid, Spain

[4] Chair in Energy Transition, School of Engineering, University of Aberdeen, Scotland

Corresponding author: Jann Michael Weinand, jann.weinand@kit.edu, +49 721 608 44444

**Abstract**

Real-world problems are becoming highly complex and, therefore, have to be solved with combinatorial optimisation (CO) techniques. Motivated by the strong increase of publications on CO, 8,393 articles from this research field are subjected to a bibliometric analysis. The corpus of literature is examined using mathematical methods and a novel algorithm for keyword analysis. In addition to the most relevant countries, organisations and authors as well as their collaborations, the most relevant CO problems, solution methods and application areas are presented. Publications on CO focus mainly on the development or enhancement of metaheuristics like genetic algorithms. The increasingly problem-oriented studies deal particularly with real-world applications within the energy sector, production sector or data management, which are of increasing relevance due to various global developments. The demonstration of global research trends in CO can support researchers in identifying the relevant issues regarding this expanding and transforming research area.

# 1. Introduction

Combinatorial optimisation (CO) has always been of great interest in the scientific community (Cacchiani et al., 2018). In CO, problems are investigated that are characterized by a finite number of possible solutions (Bjorndal et al., 1995). Whilst the discrete nature of these problems allows them to be solved in finite time by listing candidate solutions one by one and selecting the optimal solution, the number of such candidates typically grows rapidly with the input size, making many practical optimisation problems insoluble for simple enumeration schemes (Cook, 2019). Researchers in CO explore the structural features of the problems and use these features to develop both precise and approximate general solution techniques. Usually these CO problems are categorised based on their computational complexity. However, this worst-case evaluation does not always reflect the actual computational feasibility; the actual difficulty of the problems drives the development of solution approaches (Bjorndal et al., 1995). Through the development of effective methods and innovative approaches, hard real-world problems can already be solved more efficiently. At the same time, new challenges are emerging, such as the consideration of uncertain conditions, the combination of hard problems and the solving of problems in real time (Cacchiani et al., 2018).

In this context, the number of publications on CO problems has increased significantly in recent years (8,393 in 2019, cf. Section 3.1), with over 150 reviews on the subject. However, these reviews only cover certain aspects of CO: for example, there are many reviews on specific CO problems such as the quadratic assignment problem (Loiola et al., 2007), the dynamic (Pillac et al., 2013) and the multi-objective (Jozefowiez et al., 2008) vehicle routing problem, the location-routing problem (Nagy and Salhi, 2007; Prodhon and Prins, 2014) or the minimum spanning tree problem (Pop, 2020). Furthermore, many studies review metaheuristics in CO in general (Blum and Roli, 2003; Gendreau and Potvin, 2005) or in comparison with each other regarding a specific problem (e.g. traveling salesman problem (Halim and Ismail, 2019)). In addition, particular metaheuristics like ant colony optimisation (Blum, 2005), other solution algorithms like Benders decomposition (Rahmaniani et al., 2017) as well as real world applications of CO (e.g. sustainable supply chain network design (Eskandarpour et al., 2015)) are reviewed.

Hence, a review of this growing body of literature which thereby shows an overall picture of the CO research area should be beneficial for CO researchers to understand and identify research and trends in CO in its entirety. For this purpose, different quantitative and qualitative approaches are employed to understand and organise the findings of existing studies. A bibliometric analysis is one of these approaches and has the potential to provide a systematic, reproducible and transparent review process based on statistical measurements of research activities and researchers (Aria and Cuccurullo, 2017). This type of review can be classified as more objective and reliable than traditional review methods. Through a structured analysis of the extensive information on CO, the bibliometric review in this study can therefore achieve the following objectives (Aria and Cuccurullo, 2017): infer trends over time, show themes researched, identify shifts in the boundaries of the disciplines, detect most prolific authors, organisations and countries and present an overview of the extant research. However, this paper should not be read as a complete and exhaustive list of all contributions in the field of CO, but rather as an insight into the current research focus and some of the most important challenges in the field, with the bibliometric study (including all of its limitations) used as a tool to uncover them.

These objectives are achieved with the present review study on CO. For this purpose, the paper is structured as follows: Section 2 describes the main methodological aspects of this study. Subsequently, the results and discussion of the bibliometric analysis are presented in Section 3 and Section 4, respectively. Finally, Section 5 gives a summary and provides some conclusions.

## 2. Methodology

A document system together with the bibliometric characteristics of the individual documents is the object of research in bibliometric analyses. Thereby, the structure, features and patterns of the underlying science are examined using mathematical and statistical methods (Weinand, 2020). The literature database *Web of Science*[1] and the web interface *biblioshiny* of the R-tool *bibliometrix* (Aria and Cuccurullo, 2017) are used to investigate the literature on CO. In Web of Science, the *adjusted search* query in Table 1 is used. In addition to the bibliometric analysis tool bibliometrix (Section 2.1), the statistical indicators h-, g- and m-index (Section 2.2), an approach to measure trends (Section 2.3) and an algorithm for keyword and author analyses (Section 2.4) are explained in the following.

*Table 1: Search queries and resulting number of articles in the literature database Web of Science.*

| Search name | Search query | Date | Number of studies |
|---|---|---|---|
| *Optimisation in general* | (TS = "optimi*") AND LANGUAGE: (English) AND DOCUMENT TYPES: (Article) Timespan: 1990-2019. Indexes: SCI-EXPANDED, SSCI. | 23.03.2020 | 1,086,301 |
| *Initial search* | (TS = "combinatorial optimi*") Timespan: 1990-2019. Indexes: SCI-EXPANDED, SSCI. | 23.03.2020 | 8,769 |
| *Adjusted search* | (TS = "combinatorial optimi*") AND LANGUAGE: (English) AND DOCUMENT TYPES: (Article) Indexes=SCI-EXPANDED, SSCI Timespan=1990-2019 | 23.03.2020 | 8,393 |

### 2.1. R-tool bibliometrix

The main part of the analysis is based on the analysing tool on the Web of Science website and evaluations based on a newly developed algorithm (cf. Section 2.4). In addition, the R-Tool bibliometrix is applied for the examination of the corpus of literature. Bibliometrix is an open source tool for conducting comprehensive scientific mapping analyses. This tool has already been used in many bibliometric analyses. Due to its implementation in R, the package is flexible and facilitates integration with other statistical or graphical packages (Aria and Cuccurullo, 2017). One example for which bibliometrix has been used is determining the number of country collaborations (cf. Table 3).

### 2.2. Measures of influence: h-index, m-index and g-index

The h-index was introduced to facilitate quantifying the cumulative impact and relevance of an individual`s scientific output (Hirsch, 2005). Thereby an individual is associated to publications and can therefore be an author (cf. Table 1 in the Online Appendix), country (cf. Table 3), organisation (cf. Table 4) or source (cf. Table 5). The h-index reflects the number of *h* papers of an individual that have been cited at least *h* times. Together with the h-index, Hirsch (2005) also provided the m-index, which reflects the time period since the first publication of an individual by dividing the h-index by the number of years of scientific activity. Hirsch (2005) also classifies different values of the m-index with *m = 1* being a "successful scientist", *m = 2* being an "outstanding scientist" and *m = 3* being a "truly unique individual". The g-index was introduced by Egghe (2006) as an alternative of the h-index. It represents the unique largest number of the top *g* most cited articles, which together received at least $g^2$ citations. This index therefore gives a higher weighting to highly cited articles than the h-index. In the present study, however, these indices cannot be used to evaluate a single individual's scientific activity, but only for comparison with each other. This is because the publications on CO represent only a subset of an individual's total publications. As a result, for example, in relation to this subset the m-index is below one for all authors (cf. Table 1 in the Online Appendix), whereas in relation to all publications by these authors could be well above one.

---

[1] http://apps.webofknowledge.com/WOS_GeneralSearch_input.do?product=WOS&search_mode=GeneralSearch&SID=D3J AVHgH6kUCRXvVCDb&preferencesSaved=

## 2.3. Measuring trends

In some tables in Section 3, trends concerning the number of publications are indicated by means of arrow icons. To estimate the trends, publications from 2015 to 2019 are considered. The percentage increase in the number of publications per year is calculated, followed by the average from 2015 to 2019. The highest average value is then divided into five equal ranges. An example is used to show this measurement: it is assumed that the highest average annual percentage increase in publications of 20 countries between 2015 and 2019 is 100. In this case, an indication of the trend by means of arrow icons for different intervals as shown in Table 2 would be applied.

*Table 2: Arrow icons showing the trend of the increase in publications, using an example with a maximum average annual percentage increase in publications of 100 %.*

| Interval of average annual percentage increase in publications [%] | Arrow icon |
|---|---|
| [0;20] | → |
| (20;40] | ↗ |
| (40;60] | ↑ |
| (60;80] | ↑↑ |
| (80;100] | ↑↑↑ |

## 2.4. Keyword and author analysis

Keywords can also be examined with the help of bibliometrix. However, in its web interface biblioshiny, the exact strings *x* and *y* are compared with each other. If one character of a string is different from the other string, then these words are considered to be different keywords. For this reason, a separate keyword analysis algorithm was developed in MATLAB for this review. The algorithm considers similar strings as one keyword. The similarity of the strings is determined by the Levenshtein (1966) distance. The Levenshtein distance between *x* and *y* is the total cost of transforming *x* into *y* using the operations of inserting, deleting and substituting a character. Thereby, the string distance problem is equivalent to a shortest path problem defined on a graph which is constructed as follows (Spiliopoulos and Sofianopoulou, 2007):

Given the strings *x* and *y* of lengths *m* and *n*, respectively, the nodes are the points $(i, j)$ in the grid $i = 0, \ldots, m$ and $j = 0, \ldots, n$, whereby the former string is put vertically and the latter string horizontally. There are three types of directed links from the nodes *i* and *j* (Spiliopoulos and Sofianopoulou, 2007):

- vertical links $(i, j) \rightarrow (i + 1, j)$, $i = 0, \ldots, m - 1$, $j = 0, \ldots, n$, with cost 1, to represent the deletion of $x_{i+1}$,

- horizontal links $(i, j) \rightarrow (i, j + 1)$, $i = 0, \ldots, m$, $j = 0, \ldots, n - 1$, with cost 1, to represent the insertion of $y_{j+1}$ after $x_i$ (if $i = 0$, at the start of $x$),
- diagonal links $(i, j) \rightarrow (i + 1, j + 1)$, $i = 0, \ldots, m - 1$, $j = 0, \ldots, n - 1$, to represent the substitution of $x_{i+1}$ by $y_{j+1}$. There is no cost involved if $x_{i+1} = y_{j+1}$, otherwise the cost is 1.

The distance between the strings $x$ and $y$ is then given by the length of the shortest path between points (0, 0) and $(m, n)$ (Spiliopoulos and Sofianopoulou, 2007). Figure 1 shows two examples of determining Levenshtein distances in so-called dynamic programming matrices. The top rows of the matrices have zero values, since the insertions at the start of the vertical string are not penalised. After filling in the matrices according to the above-mentioned links, the minimum value in the last row shows the distance. In the first example in Figure 1a there are three substitutions (P-U and E with B-I and A) and six insertions (N and O-R-I-A-L), whereas in the second example there are only six substitutions (P-U-T with B-I-N and I-O-N with O-R-I, cf. Figure 1b).

a)

|   |   | C | O | M | B | I | N | A | T | O | R | I | A | L |
|---|---|---|---|---|---|---|---|---|---|---|---|---|---|---|
|   | 0 | 1 | 2 | 3 | 4 | 5 | 6 | 7 | 8 | 9 | 10 | 11 | 12 | 13 |
| C | 1 | 0 | 1 | 2 | 3 | 4 | 5 | 6 | 7 | 8 | 9 | 10 | 11 | 12 |
| O | 2 | 1 | 0 | 1 | 2 | 3 | 4 | 5 | 6 | 7 | 8 | 9 | 10 | 11 |
| M | 3 | 2 | 1 | 0 | 1 | 2 | 3 | 4 | 5 | 6 | 7 | 8 | 9 | 10 |
| P | 4 | 3 | 2 | 1 | 1 | 2 | 3 | 4 | 5 | 6 | 7 | 8 | 9 | 10 |
| U | 5 | 4 | 3 | 2 | 2 | 2 | 3 | 4 | 5 | 6 | 7 | 8 | 9 | 10 |
| T | 6 | 5 | 4 | 3 | 3 | 3 | 3 | 4 | 4 | 5 | 6 | 7 | 8 | 9 |
| E | 7 | 6 | 5 | 4 | 4 | 4 | 4 | 4 | 5 | 5 | 6 | 7 | 8 | 9 |

b)

|   |   | C | O | M | B | I | N | A | T | O | R | I | A | L |
|---|---|---|---|---|---|---|---|---|---|---|---|---|---|---|
|   | 0 | 1 | 2 | 3 | 4 | 5 | 6 | 7 | 8 | 9 | 10 | 11 | 12 | 13 |
| C | 1 | 0 | 1 | 2 | 3 | 4 | 5 | 6 | 7 | 8 | 9 | 10 | 11 | 12 |
| O | 2 | 1 | 0 | 1 | 2 | 3 | 4 | 5 | 6 | 7 | 8 | 9 | 10 | 11 |
| M | 3 | 2 | 1 | 0 | 1 | 2 | 3 | 4 | 5 | 6 | 7 | 8 | 9 | 10 |
| P | 4 | 3 | 2 | 1 | 1 | 2 | 3 | 4 | 5 | 6 | 7 | 8 | 9 | 10 |
| U | 5 | 4 | 3 | 2 | 2 | 2 | 3 | 4 | 5 | 6 | 7 | 8 | 9 | 10 |
| T | 6 | 5 | 4 | 3 | 3 | 3 | 3 | 4 | 4 | 5 | 6 | 7 | 8 | 9 |
| A | 7 | 6 | 5 | 4 | 4 | 4 | 3 | 4 | 5 | 6 | 7 | 7 | 8 |
| T | 8 | 7 | 6 | 5 | 5 | 5 | 5 | 4 | 3 | 4 | 5 | 6 | 7 | 8 |
| I | 9 | 8 | 7 | 6 | 6 | 5 | 6 | 5 | 4 | 4 | 5 | 5 | 6 | 7 |
| O | 10 | 9 | 8 | 7 | 7 | 6 | 6 | 6 | 5 | 4 | 5 | 6 | 6 | 7 |
| N | 11 | 10 | 9 | 8 | 8 | 7 | 6 | 7 | 6 | 5 | 5 | 6 | 7 | 7 |
| A | 12 | 11 | 10 | 9 | 9 | 8 | 7 | 6 | 7 | 6 | 6 | 6 | 6 | 7 |
| L | 13 | 12 | 11 | 10 | 10 | 9 | 8 | 7 | 7 | 7 | 7 | 7 | 7 | 6 |

*Figure 1: Dynamic programming matrices for determining the Levenshtein distance between a) "compute" and "combinatorial" and b) "computational" and "combinatorial".*

In the developed algorithm, the Levenshtein distance is used to match equivalent strings under one keyword. This way the number of occurrences of the keywords can be determined more accurately. Strings of five or more characters are grouped for a Levenshtein distance up to one and strings of nine or more characters are grouped for a Levenshtein distance up to two. The latter case should ensure that, for example, the plural of a word and the simultaneous use of a hyphen is recognized as the same keyword (e.g. *metaheuristic* and *meta-heuristics*). Besides the number of occurrences of the keywords, the algorithm also determines the mean publication year and the mean citations of all articles with the respective keyword. In addition, the algorithm examines the keywords with regard to their simultaneous occurrence in the same articles (cf. Section 3.4). This latter function is also used to identify the collaboration of authors (cf. Figure 1 in the Online Appendix). The MATLAB script is applicable to any other bibliometric analysis and can be provided upon request.

## 3. Results

In the following, the main characteristics of the research field on CO are presented (cf. Section 3.1). Afterwards, an overview of contributions and collaborations of different countries and organisations is given in Section 3.2. Subsequently, the most relevant sources and studies are highlighted (cf. Section 3.3) before the most relevant topics on CO are discussed (cf. Section 3.4).

## 3.1. Development of the research field

The 8,393 articles have been published in 1,415 different sources. A total of 14,423 authors were involved in the articles and the average number of citations per document is 23.55. This high citation rate is due to some highly cited publications (cf. Section 3.3); 55% of the articles are cited less than ten times. The number of publications per year on CO has increased quite steadily over the years from 13 in 1990 to 533 in 2019. This could be related to the general increase of publications in the field of operations research. However, when comparing the share of articles on CO with the total number of articles on optimisation in general (search query *optimisation in general* in Table 1), a slight increase can be observed: in 1990, 15,868 articles on optimisation in general and 76 (0.5%) on CO had already appeared, and in 2019, the share is 0.8% with 8,456 in the total number of 1,100,191 publications.

## 3.2. Publication distribution and collaboration of countries, organisations and authors

In total, authors from 85 countries have contributed to articles on CO. The fact that CO is only extensively researched in a limited number of countries is shown by the share of the top 20 relevant countries in Table 3, which, with a number of 7,157 articles, are involved in 85% of the publications on CO. In the top 20 are Australia, eleven countries from Europe, five from Asia and three from America. By far the most articles were authored by researchers from the USA (23%), followed by China (14%), France (8%), Germany (7%) and Japan (7%). The development of the number of articles in these top five relevant countries in Figure 2a is of interest: while authors in the USA published the most articles each year from 1990 to 2013, China was responsible for the fewest publications in the 1990s. However, the number of annual articles by Chinese authors has been increasing almost exponentially since then, and since 2014 most of the annual articles on CO are written by Chinese authors. Furthermore, as the arrow icons in Table 3 show, the annual percentage increase in the number of articles since 2015 among the top 20 countries is highest in China (cf. Section 2.3). This could be related to the rapid growth of the Chinese economy and the associated increase in energy consumption (Zhang et al., 2017). As Section 3.4.3 shows, the production sector and the energy sector are the most important application areas of CO research.

However, these numbers of published articles do not provide any indication of the importance of CO research in the overall research of a country. For this reason, the total publications of a country are also shown in Table 3 in relation to the mean annual (between 2012 and 2018) gross domestic expenditure on R&D of the respective country (UNESCO, 2020). In the top five most productive countries, with the exception of France, CO seems to be of comparatively low importance. In contrast, CO research seems to be of great priority in Portugal, Iran, Spain, Belgium and Canada. In addition, the USA has by far the highest number of citations (67,298) and the highest h- (104), g- (212) and m-index (3.47). However, in terms of average article citations, the USA is only in sixth position (35.1), behind Australia (41.5), Belgium (40.9), Canada (40.8), England (36.8) and Switzerland (35.2). Due to the high number of citations, the g-index in these countries is also comparatively high. In relation to the number of publications, the m-index is high in some countries, such as Iran, which published its first articles on CO in 2001.

Table 3: The top 20 of the most productive countries in terms of publications on combinatorial optimisation. The numbers for "total publications" do not have to sum up to 8,393 or 100%, since more than one country could have contributed to a single publication. The percentages for "total publications" and "corresponding author's country" refer to the total number of 8,393 articles, while for "single publication" and "collaborative publication" they refer to the number of publications of "corresponding author's country".

| Country | Total publications | | Trend | h-index | g-index | m-index | TPGD[2] (PPP$-billion)$^{-1}$ | Average article citations | Corresponding author's country | | Single country publication | | Collaborative publications | |
|---|---|---|---|---|---|---|---|---|---|---|---|---|---|---|
| | No. | % | | | | | | No. | No. | % | No. | % | No. | % |
| USA | 1,918 | 23 | ↗ | 104 | 212 | 3.47 | 4 | 35.1 | 1,388 | 17 | 977 | 70 | 411 | 30 |
| China | 1,197 | 14 | ↑↑↑ | 62 | 103 | 2.21 | 3 | 16.6 | 1,068 | 13 | 812 | 76 | 256 | 24 |
| France | 676 | 8 | ↑ | 53 | 86 | 1.77 | 11 | 19.2 | 499 | 6 | 319 | 64 | 180 | 36 |
| Germany | 586 | 7 | ↑ | 49 | 76 | 1.63 | 5 | 17.4 | 432 | 5 | 279 | 65 | 153 | 35 |
| Japan | 547 | 7 | ↗ | 48 | 86 | 1.60 | 3 | 18.4 | 495 | 6 | 376 | 76 | 119 | 24 |
| England | 486 | 6 | ↑ | 55 | 123 | 1.83 | 11 | 36.8 | 330 | 4 | 205 | 62 | 125 | 38 |
| Italy | 425 | 5 | ↗ | 42 | 109 | 1.45 | 14 | 32.5 | 325 | 4 | 236 | 73 | 89 | 27 |
| Canada | 415 | 5 | ↑↑ | 52 | 124 | 1.79 | 15 | 40.8 | 254 | 3 | 162 | 64 | 92 | 36 |
| Spain | 398 | 5 | ↑↑ | 36 | 68 | 1.20 | 20 | 17.7 | 304 | 4 | 204 | 67 | 100 | 33 |
| Brazil | 308 | 4 | ↑↑ | 36 | 69 | 1.29 | 8 | 19.7 | 256 | 3 | 192 | 75 | 64 | 25 |
| Australia | 251 | 3 | ↑↑ | 36 | 100 | 1.24 | 11 | 41.5 | 166 | 2 | 115 | 69 | 51 | 31 |
| India | 251 | 3 | ↑↑ | 30 | 48 | 1.03 | 4 | 13.5 | 214 | 3 | 170 | 79 | 44 | 21 |
| Netherlands | 215 | 3 | ↗ | 31 | 60 | 1.07 | 13 | 20.3 | 130 | 2 | 82 | 63 | 48 | 37 |
| Taiwan | 210 | 3 | → | 42 | 67 | 1.45 | n.a. | 25.3 | 185 | 2 | 158 | 85 | 27 | 15 |
| Belgium | 199 | 2 | ↑ | 34 | 88 | 1.36 | 15 | 40.9 | 149 | 2 | 82 | 55 | 67 | 45 |
| Turkey | 179 | 2 | ↑↑ | 32 | 49 | 1.14 | 11 | 18.2 | 151 | 2 | 120 | 79 | 31 | 21 |
| Austria | 171 | 2 | ↗ | 32 | 60 | 1.10 | 13 | 24.1 | 120 | 1 | 72 | 60 | 48 | 40 |
| Iran | 163 | 2 | ↑↑ | 30 | 45 | 1.58 | 23 | 17.0 | 148 | 2 | 130 | 88 | 18 | 12 |
| Portugal | 155 | 2 | ↑ | 26 | 51 | 1.04 | 39 | 20.0 | 107 | 1 | 74 | 69 | 33 | 31 |
| Switzerland | 153 | 2 | ↑ | 32 | 72 | 1.10 | 10 | 35.2 | 108 | 1 | 56 | 52 | 52 | 48 |

Most collaborative publications have been produced by authors from USA and China (146), USA and Canada (73), USA and Germany (69), USA and France (60) as well as England and China (55). By using the corresponding author for assigning the articles to a country (cf. Table 3), the countries which frequently participate in international cooperation can be identified. Switzerland (48%) has the largest share of collaborative publications, ahead of Belgium (45%) and Austria (40%). In contrast, authors from Iran (88% single country publications) and Taiwan (85%) show the lowest share of collaborative publications. In general, the share of international collaboration or intranational collaboration is rather low, which is also reflected by the number of single-authored publications (1.087; 12%).

A total of 3,539 different organisations from the 85 countries were participating in the publications on CO. The top 20 of the most relevant organisations are listed in Table 4. These organisations have been involved in 1,696 publications (20%) so far. Most of the articles have originated from authors of the top five organisations *French National Centre for Scientific Research* (253, France), *University of California System* (162, USA), *Chinese Academy of Sciences* (118, China), *National Institute for Research in Digital Science and Technology* (115, France) and *Massachusetts Institute of Technology* (101, USA). However, it should be noted that some of these organisations are associations of several institutions, such as the *Chinese Academy of Sciences* or the *University of California System*. The annual publication volume of these five organisations in Figure 2b shows a slightly rising trend in addition to annual fluctuations. The trend in the annual percentage increase in publications (cf. Table 4) shows a particularly strong increase for *Chinese Academy of Sciences*, *University System of Georgia*, *University of Tokyo* and *Huazhong University of Science and Technology*. The publications on CO from the

---

[2] TPGD = Total publications (TP) per gross domestic (GD) expenditure on R&D (UNESCO (2020)) expressed in the Purchasing Power Parity (PPP) of the respective country. As gross domestic expenditure on R&D, the mean annual value between 2012 and 2018 is used.

*University Of California System* have the highest h-index (37), followed by the *French National Centre for Scientific Research* (31) and the *Massachusetts Institute of Technology* (30). However, the order changes when the number of citations is taken into account for the most cited articles: for the g-index, the *Universite Libre de Bruxelles* (75) is at the top of the list, ahead of the *Massachusetts Institute of Technology* (70) and the *University of Montreal* (67). The publications of the *University of California System* and *Huazhong University of Science and Technology* are particularly relevant with regard to the first publication year, with m-indexes of 1.28 and 1.25 respectively. The majority of collaborations took place between organisations of the same country (cf. Figure 3): most collaborations were conducted by the *University of Montreal* and *Polytechnique Montreal* (41), the *French National Centre for Scientific Research* and the *National Institute for Research in Digital Science and Technology* (38) as well as the *National Institute for Research in Digital Science and Technology* and the *University of Lille* (31). The publication distribution and collaboration of the most relevant authors can be found in the Online Appendix.

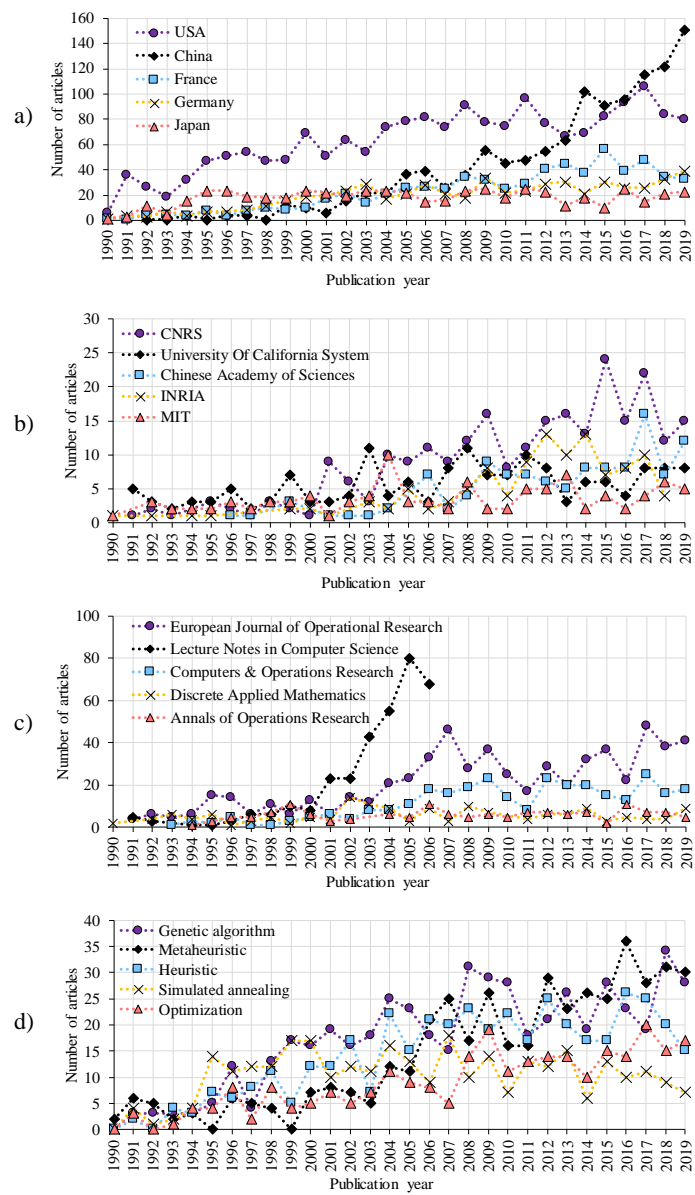

*Figure 2: Annual development of publications of the top five most productive countries a), institutes b), sources c) and keywords d). For reasons of clarity, zero values are not displayed. Web of Science stopped covering articles from the source Lecture Notes in Computer Science in 2007.*

*Table 4: The top 20 of the most productive organisations in terms of publications on combinatorial optimisation. The percentage values refer to the total of 8,393 publications.*

| Organisation | Country | Total publications No. | % | Trend | h-index | g-index | m-index |
|---|---|---|---|---|---|---|---|
| French National Centre for Scientific Research CNRS | France | 253 | 3 | ↑ | 31 | 52 | 1.07 |
| University of California System | USA | 162 | 2 | ↑ | 37 | 62 | 1.28 |
| Chinese Academy of Sciences | China | 118 | 1 | ↑↑↑ | 20 | 41 | 0.83 |
| National Institute for Research in Digital Science and Technology INRIA | France | 115 | 1 | ↑ | 25 | 39 | 0.83 |
| Massachusetts Institute of Technology MIT | USA | 101 | 1 | ↗ | 30 | 70 | 1.00 |
| State University System of Florida | USA | 101 | 1 | ↑ | 25 | 41 | 1.04 |
| University of Montreal | Canada | 89 | 1 | ↑ | 27 | 67 | 0.93 |
| Universite Libre de Bruxelles | Belgium | 88 | 1 | ↑ | 22 | 75 | 0.88 |
| University System of Georgia | USA | 84 | 1 | ↑↑↑ | 25 | 53 | 0.89 |
| University of Bologna | Italy | 80 | 1 | ↑ | 22 | 46 | 0.88 |
| University of Tokyo | Japan | 78 | 1 | ↑↑↑ | 21 | 41 | 0.84 |
| Carnegie Mellon University | USA | 65 | 1 | ↑ | 23 | 48 | 0.79 |
| Tsinghua University | China | 64 | 1 | ↑ | 19 | 40 | 0.73 |
| University of Texas System | USA | 64 | 1 | ↑ | 17 | 45 | 0.63 |
| Huazhong University of Science and Technology | China | 63 | 1 | ↑↑↑ | 25 | 41 | 1.25 |
| Indian Institute of Technology System IIT System | India | 63 | 1 | ↑↑ | 15 | 29 | 0.54 |
| Nanyang Technological University | Singapore | 63 | 1 | ↑ | 27 | 41 | 1.08 |
| Russian Academy of Sciences | Russia | 63 | 1 | ↑↑ | 9 | 16 | 0.31 |
| Polytechnic University of Catalonia | Spain | 62 | 1 | ↑↑ | 19 | 47 | 0.79 |
| Universidade de Lisboa | Portugal | 62 | 1 | ↑↑ | 17 | 39 | 0.68 |

## 3.3. Most relevant sources and articles

The articles on CO have been published in 1,415 different sources. A large proportion of the articles on CO (35%) is published in the top 20 most productive sources (cf. Table 5) *European Journal of Operational Research* (EJOR), *Lecture Notes in Computer Science* (LNCS), *Computers & Operations Research*, *Discrete Applied Mathematics* and *Annals of Operations Research*. Most articles on CO were published in EJOR, with 7% of all publications. Therefore, it is not surprisingly that this source has the highest h- (58) and g-index (99), followed by *Computers & Operations Research* regarding the h-index (46) and by *Lecture Notes in Computer Science* regarding the g-index (87). The source *Expert Systems with Applications* has a particularly high m-index (2.07) due to a high h-index (31) in relation to the starting year of the first publication on CO (2005). In the most relevant source, EJOR, the share of articles on CO in all publications has increased steadily, from 2% in 1990 to 6% in 2019. In addition to the findings from Section 3.1, this again demonstrates the increasing importance of CO in the research field of operations research.

However, when examining the development of the annual publications of the top five most relevant sources in Figure 2c, it is striking that the annual publications of LNCS increased exponentially between 2001 and 2005. If this trend had continued, LNCS would have been in first rank among the most productive sources on CO. However, from 2007 onwards, no publications by LNCS are listed in Web of Science anymore. A reason for this could be that LNCS publishes mostly conference proceedings. The trend in the annual percentage increase in publications shows a strong upward trend for only a few sources, with *Applied Soft Computing* at the top (cf. Table 5).

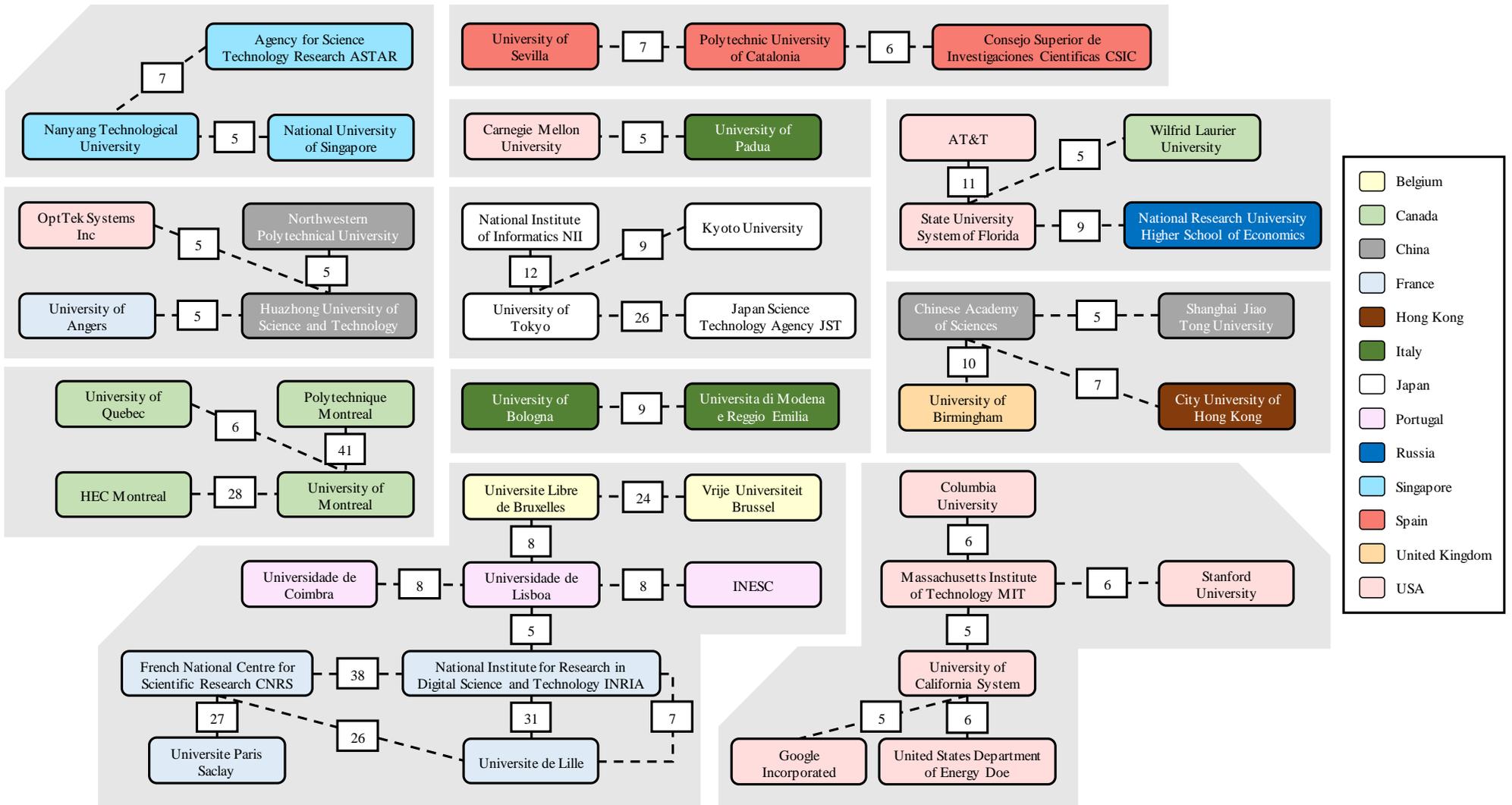

Figure 3: Collaboration network of the top 20 most productive organisations together with their top three most productive collaborations. Collaborations with less than five collaborative publications are not shown.

Table 2 in the Online Appendix lists the 20 globally most cited articles in the CO research area. Many of these articles are about newly developed metaheuristics like ant colony optimisation (Dorigo et al., 1996; Dorigo and Gambardella, 1997), Harmony Search (Geem et al., 2001) or Variable Neighbourhood Search (Mladenović and Hansen, 1997). The study by Dorigo et al. (1996) on ant colony optimisation is cited most frequently (5,646). Furthermore, there are some articles in the top 20 that deal with the development of new algorithms (evolutionary algorithms (Han and Kim, 2002; Yao et al., 1999), approximation algorithms (Goemans and Williamson, 1995)) or the enhancement of existing metaheuristics (MAX-MIN ant system (Stützle and Hoos, 2000)). It is noticeable that the newly developed metaheuristics are mostly tested and benchmarked using the Traveling Salesman problem (TSP) (Dorigo et al., 1996; Dorigo and Gambardella, 1997; Geem et al., 2001; Mladenović and Hansen, 1997; Stützle and Hoos, 2000). The studies described in this paragraph also have by far the most local citations, i.e. within the 8,393 articles examined here. This means that these fundamental new methods are either applied or extended in many other articles. Ant colony optimisation is the most common method with four studies in the top 20 of the most cited articles (Dorigo et al., 1996; Dorigo and Blum, 2005; Dorigo and Gambardella, 1997; Stützle and Hoos, 2000).

In addition to the development of these theoretical approaches, CO methods are applied in case studies in the top 20 articles, such as vision / image processing (Boykov and Funka-Lea, 2006; Boykov and Kolmogorov, 2004; Jaqaman et al., 2008), quantum computers (Knill et al., 2001) or atomic decomposition (Donoho and Huo, 2001). In contrast to the studies on developments of metaheuristics and algorithms described above, the local citations of these articles on specific applications are very low. Furthermore, only two of the top 20 most cited articles were published in the top five most relevant sources: one each in *European Journal of Operational Research* and *Computers & Operations Research* (cf. Table 2 in the Online Appendix).

*Table 5: The top 20 of the most productive scientific sources in terms of publications on combinatorial optimisation. The percentage values refer to the total of 8,393 publications. The trend in number of publications is not shown for the Lecture Notes in Computer Science, since this source is not indexed in Web of Science since 2007.*

| Source | Total publications (TP) No. | % | Trend | Average article citations | h-index | g-index | m-index |
|---|---|---|---|---|---|---|---|
| European Journal of Operational Research | 615 | 7 | ↑ | 26 | 58 | 99 | 2.00 |
| Lecture Notes in Computer Science | 335 | 4 | - | 29 | 28 | 87 | 0.97 |
| Computers & Operations Research | 306 | 4 | ↗ | 8 | 46 | 44 | 1.70 |
| Discrete Applied Mathematics | 158 | 2 | ↗ | 25 | 23 | 58 | 0.77 |
| Annals of Operations Research | 149 | 2 | ↗ | 8 | 31 | 31 | 1.19 |
| Journal of Combinatorial Optimization | 138 | 2 | ↑↑ | 24 | 17 | 53 | 0.74 |
| Mathematical Programming | 125 | 1 | → | 18 | 30 | 43 | 1.03 |
| Applied Soft Computing | 120 | 1 | ↑↑↑ | 24 | 25 | 46 | 1.56 |
| Expert Systems with Applications | 120 | 1 | ↗ | 24 | 31 | 51 | 2.07 |
| Computers & Industrial Engineering | 110 | 1 | ↑↑ | 24 | 25 | 48 | 0.93 |
| Journal of Heuristics | 108 | 1 | ↗ | 22 | 30 | 47 | 1.36 |
| Theoretical Computer Science | 87 | 1 | ↑ | 17 | 19 | 33 | 0.73 |
| International Journal of Production Research | 82 | 1 | ↑↑ | 14 | 23 | 30 | 0.82 |
| Operations Research Letters | 79 | 1 | ↗ | 21 | 18 | 39 | 0.62 |
| Journal of the Operational Research Society | 76 | 1 | ↗ | 28 | 21 | 46 | 0.72 |
| Journal of Global Optimization | 70 | 1 | ↗ | 7 | 18 | 19 | 0.69 |
| Discrete Optimization | 66 | 1 | ↗ | 31 | 13 | 44 | 0.93 |
| Information Sciences | 65 | 1 | ↑ | 13 | 22 | 27 | 0.85 |
| Algorithmica | 61 | 1 | ↑ | 21 | 13 | 33 | 0.45 |
| Computational Optimization and Applications | 60 | 1 | ↗ | 11 | 19 | 23 | 0.83 |

However, the most cited articles described above do not provide a complete picture of research trends in CO, as many are relatively old with the most recent article dating from 2008. Therefore, Table 6 shows the top ten articles with the highest annual citation rates that have appeared since 2010. With 108 citations, Deng et al. (2019) is cited the most annually. Similar to the sixth most cited study by Nouiri et al. (2018), a new particle swarm optimisation method is developed in Deng et al. (2019). An advanced method, in this case a search algorithm, is also introduced in Wen and Yin (2013). Some other studies are more application-oriented on phosphorus systems for white light emitting diodes (Xia et al., 2016), biological transport networks (Tero et al., 2010) or fatty acids production (Xu et al., 2013).

Furthermore, three survey studies are included in Table 6: firstly, on hyper heuristics (Burke et al., 2013), i.e. heuristics that choose an appropriate solution heuristic depending on the problem. The other two publications are strongly linked regarding their content. Sörensen (2015) discusses the fact that many metaheuristics have been developed in recent years and that these are largely based on metaphors about natural or man-made processes, for example ant or bee colony optimisation. The author argues that besides some innovative studies of high quality, many papers have been published that are justified only by the fact that the developed metaheuristics are based on novel metaphors. Blum et al. (2011) also motivate their study on hybrid metaheuristics by the fact that research on metaheuristics for CO problems was mostly algorithm-oriented in the past. Nevertheless, the authors see a trend that the focus of research on CO metaheuristics is shifting from this algorithm orientation to a problem orientation. As a result, metaheuristics are now often hybridized with other optimisation techniques in order to find the best approaches to solving problems.

*Table 6: Articles among the scientific contributions on combinatorial optimisation, which have the highest annual citation rate since 2010.*

| Article title | Global citations | | Publication year | Source |
|---|---|---|---|---|
| | *No.* | *Per year* | | |
| *A novel intelligent diagnosis method using optimal LS-SVM with improved PSO algorithm (Deng et al., 2019)* | 108 | 108 | 2019 | Soft Computing |
| *Recent developments in the new inorganic solid-state LED phosphors (Xia et al., 2016)* | 264 | 66 | 2016 | Dalton Transactions |
| *Quantum annealing with manufactured spins (Johnson et al., 2011)* | 533 | 59 | 2011 | Nature |
| *Hyper-heuristics: a survey of the state of the art (Burke et al., 2013)* | 354 | 51 | 2013 | Journal of the Operational Research Society |
| *Metaheuristics-the metaphor exposed (Sörensen, 2015)* | 223 | 45 | 2015 | International Transactions in Operational Research |
| *An effective and distributed particle swarm optimization algorithm for flexible job-shop scheduling problem (Nouiri et al., 2018)* | 82 | 41 | 2018 | Journal of Intelligent Manufacturing |
| *Rules for biologically inspired adaptive network design (Tero et al., 2010)* | 375 | 38 | 2010 | Science |
| *Modular optimization of multi-gene pathways for fatty acids production in E. coli (Xu et al., 2013)* | 262 | 37 | 2013 | Nature Communications |
| *Hybrid metaheuristics in combinatorial optimization: a survey (Blum et al., 2011)* | 323 | 36 | 2011 | Applied Soft Computing |
| *A feasible method for optimization with orthogonality constraints (Wen and Yin, 2013)* | 225 | 32 | 2013 | Mathematical Programming |

As already described, the article by Dorigo et al. (1996) has the most global citations (5,646) among the articles examined here and is also very often local cited (235) by these 8,393 studies. However, five studies are even more frequently cited locally by the 8,393 articles (cf. Table 7). The focus of the most local cited study is on the identification and handling of NP-complete (combinatorial optimisation) problems, i.e. problems that can be solved in polynomial time on a non-deterministic Turing Machine (Garey and Johnson, ca. 2009). Three other publications in Table 7 deal with the introduction of new metaheuristics (Simulated Annealing (Kirkpatrick et al., 1983) and Tabu Search (Glover, 1989)) or a review of a metaheuristic (Genetic algorithm (Goldberg, 2012)). These three metaheuristics are also the most frequently employed in the 8,393 studies on CO (cf. Section 3.4).

*Table 7: Five most frequently cited references in the articles on combinatorial optimisation.*

| Article title | Local citations | Publication year | Source |
|---|---|---|---|
| Computers and intractability: A guide to the theory of NP-completeness (Garey and Johnson, ca. 2009) | 911 | 1979 | A series of books in the mathematical sciences |
| Optimization by Simulated Annealing (Kirkpatrick et al., 1983) | 702 | 1983 | Science |
| Genetic algorithms in search, optimization, and machine learning (Goldberg, 2012) | 527 | 1989 | - |
| Tabu Search—Part I (Glover, 1989) | 276 | 1989 | ORSA Journal on Computing |
| "Neural" computation of decisions in optimization problems (Hopfield and Tank, 1985) | 266 | 1985 | Biological cybernetics |

### 3.4. Most relevant subjects

In this section, firstly the algorithm based on the Levenshtein distance is evaluated (cf. Section 3.4.1). Secondly, the most relevant keywords and thus topics in the research area of CO are shown in Section 3.4.2. Section 3.4.3 then attempts to determine the most relevant application areas of CO.

### 3.4.1. Evaluation of keyword algorithm

The developed algorithm led to a much better recognition of related keywords than was the case with the web interface *biblioshiny* of the R-tool bibliometrix. For example, different spellings with or without hyphen or in singular or plural are now combined. The keyword *combinatorial optimization*, which occurs most frequently in the articles with 3.051 times, serves as an example. The different spellings that have been grouped for this keyword are the following, whereby the change in spelling compared to the first keyword is shown in bold: *combinatorial optimization*, *combinatorial optimization**S***, *combinatorial optimi**S**ation*, *combinatorial optimi**S**ation**S*** and *combinatori**C** optimization*. Thus 7% more appearances of the keyword *combinatorial optimization* could be identified (cf. Figure 4). For the seven most relevant keywords shown in Figure 4, the greatest improvement was achieved for the keyword *genetic algorithm*: here the recognition rate increased by 99%. However, there are also examples that seem to be always stated in the same way: for *simulated annealing*, the developed algorithm yielded no improvement.

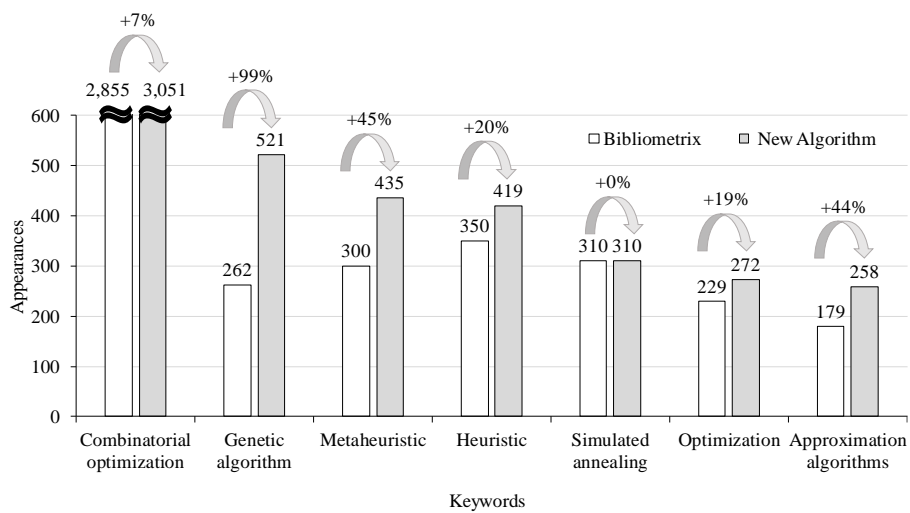

*Figure 4: Comparison of the appearance of the most relevant keywords determined on the one hand with the R-tool bibliometrix and on the other hand with the new algorithm developed for this study.*

### 3.4.2. Combinatorial optimisation problems and solution methods

In this section, the most relevant keywords on CO that appear in at least 1% of publications are discussed. As already shown in Section 3.4.1, the keyword *combinatorial optimization* occurs most frequently (in 3051 publications). This is not surprising since the keyword corresponds to the overall topic of this bibliometric analysis and thus the search query in Web of Science. Therefore, the keyword *combinatorial optimization* is excluded in the following analysis. Thus, the top five relevant keywords are *genetic algorithm* (521 occurrences), *metaheuristic* (435), *heuristic* (419), *simulated annealing* (310) and *optimization* (272). The annual occurrence of these keywords fluctuates, but in general there is an increasing trend (except for simulated annealing, cf. Figure 2d). The fact that simulated annealing is less frequently covered in actual CO publications is further demonstrated by the low mean publication year (2005.7, cf. Figure 5). In Figure 5, the 25 keywords that appear in at least 1% of publications are divided into different categories and their mean publication year is shown. In addition, the correlation matrix in Figure 6, which shows how often these keywords occur together in articles, is also relevant for the following analysis.

The most frequently addressed specific optimisation problems are the TSP (225 appearances), integer programming (225) and the scheduling problem (219). While no clear trend can be identified for integer programming, the TSP is mainly investigated in connection with the metaheuristics ant colony optimisation (20), genetic algorithms (18) as well as local search (16) and the scheduling problem with tabu search (13) and simulated annealing (10) (cf. Figure 6). The TSP is examined so frequently since it is representative of CO problems. If an efficient (polynomial-time) algorithm could be found for the TSP, then efficient algorithms could also be found for all other NP-complete problems (Hoffman et al., 2013). Integer programming problems (e.g. assignment problem) are also often closely related to CO (Conforti et al., 2014). However, not every CO problem can be formulated as an integer programming problem, if its feasible region is infinite (Ibaraki, 1976). The scheduling problem is often used for decision support, for example in the case of project scheduling in project management (Hartmann and Briskorn, 2010), personnel scheduling (van den Bergh et al., 2013) or maintenance scheduling (Froger et al., 2016).

The keywords with the highest mean publication year represent the most recent topics in CO. The two most recent of the specific optimisation problems are robust optimisation (mean publication year: 2014.0) and multi-objective optimisation (2011.4). This is also demonstrated by the trends in the annual percentage increase in publications on these topics since 2015. The difficulties, that several objectives have to be optimised simultaneously and that not all parameters are known in advance, are often encountered when applying optimisation techniques to real-world problems (Schmidt et al., 2019). The method of robust optimisation includes several approaches to protect a decision maker against parameter ambiguity and stochastic uncertainty. Thereby, the manager must determine what it means for him to have a robust solution. Based on worst-case analysis, a solution is evaluated using the realization of the most unfavourable uncertainty (Gabrel et al., 2014). Multi-objective optimisation involves optimising multiple objectives at the same time by selecting a (Pareto) efficient solution that cannot be improved in one objective without worsening it in another objective. In recent years, the concepts of both areas have been combined into multi-objective robust optimisation (Schmidt et al., 2019).

The most prominent general solution techniques for solving these CO problems are metaheuristics (435 appearances), heuristics (419), optimisations (272) and approximation algorithms (258) (cf. Figure 5). Real-world CO problems are usually large and exact solution procedures are mostly inadequate. Hence heuristics are mainly used in practice to solve complex CO problems (Hertz and Widmer, 2003). In the past, typically specialized heuristics were developed. However, this approach changed over the years: more general (metaheuristics) and less specialized solution approaches emerged. The motivation here is that applying a metaheuristic to a specific problem or problem class requires less effort than developing a specialized heuristic from scratch (Gendreau and Potvin, 2005). For heuristics and metaheuristics, for certain inputs good solutions (i.e. close to the optimal solution of a problem) are determined, but it is often uncertain why the heuristics work well. In this context, approximation algorithms are helpful, which bring mathematical rigor to the study of heuristics. Thus it can be proven how well a heuristic performs on all instances and an idea of the types of instances on which a heuristic does not perform well can be given (Williamson and Shmoys, 2011).

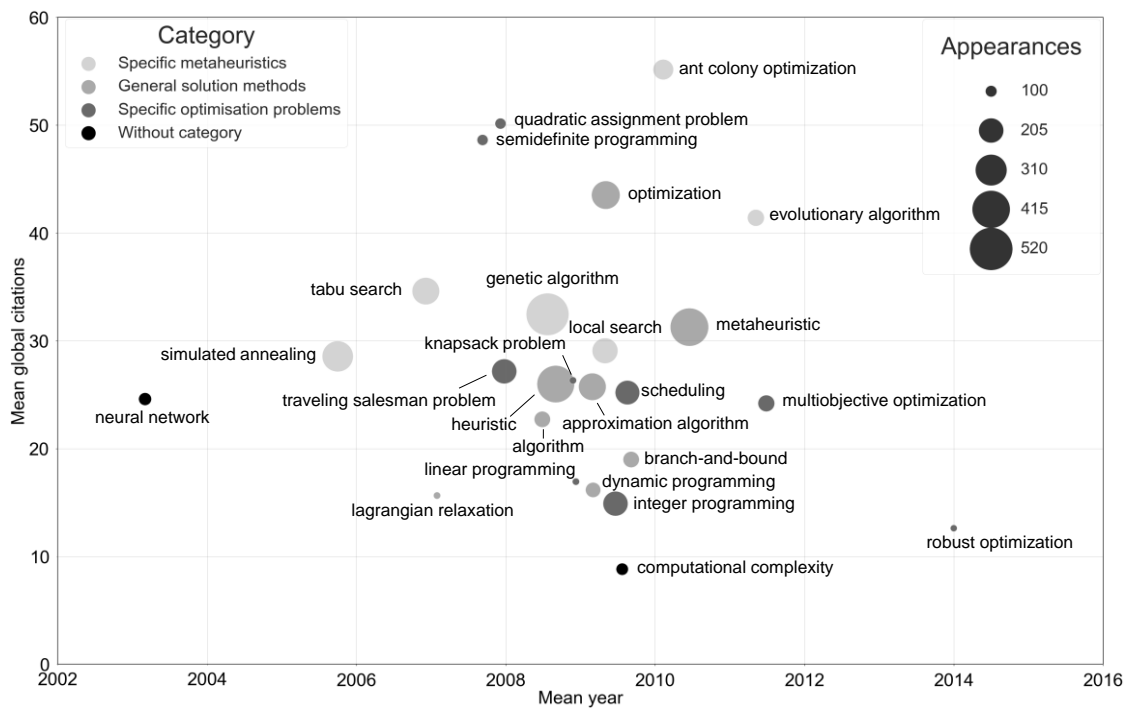

*Figure 5: The top 25 most relevant keywords on CO with their number of appearances, mean global citations and mean publication year in the 8,393 articles.*

Furthermore, many different specific metaheuristics are included in the 8,393 studies (cf. Figure 5). Various taxonomies can be found in the literature to distinguish metaheuristics (see Zäpfel et al. (2010)). To give one example: Hertz and Widmer (2003) distinguish metaheuristics by two principles, namely local search and population search. The authors define the local search methods as an intensive exploration of the solution space by moving from the current solution to another promising solution in the neighbourhood at each iteration. In comparison, the population search consists of maintaining a variety of good solutions and combining them to produce better solutions. The three metaheuristics most frequently found in the 8,393 studies on CO are among the classic examples of population search and local search: Genetic algorithm (521 occurrences; population search) as well as Simulated Annealing (310; local search) and Tabu search (257; local search). These three metaheuristics are also often considered jointly in studies having at least 34 shared occurrences (c.f. Figure 6). The ant colony optimisation, which is so frequently covered in the top 20 most cited articles, is also among the most relevant specific metaheuristics with 175 appearances. The articles on ant colony optimisation are most cited on average (55, cf. Figure 5) followed by articles on particle swarm optimisation (52) and articles on quadratic assignment problems (50). For more information on the chronological development of the research field of metaheuristics please refer to Sörensen et al. (2018). Information on the most important topics of countries, organisations and sources can be found in the Online Appendix.

| A | B | C | D | E | F | G | H | I | J | K | L | M | N | O | P | Q | R | S | T | U | V | W | X | Y | Z |   |
|---|---|---|---|---|---|---|---|---|---|---|---|---|---|---|---|---|---|---|---|---|---|---|---|---|---|---|
| 258 | 0 | 2 | 1 | 2 | 11 | 8 | 10 | 2 | 6 | 3 | 2 | 1 | 1 | 0 | 0 | 7 | 1 | 1 | 0 | 10 | 8 | 10 | 9 | 1 | 4 | Approximation Algorithm (A) |
|  | 175 | 1 | 21 | 2 | 0 | 1 | 0 | 4 | 3 | 9 | 7 | 22 | 9 | 11 | 4 | 0 | 4 | 1 | 0 | 0 | 20 | 0 | 0 | 1 | 1 | Ant Colony Optimization (B) |
|  |  | 141 | 12 | 3 | 0 | 1 | 9 | 6 | 10 | 13 | 5 | 12 | 3 | 2 | 3 | 1 | 0 | 0 | 0 | 0 | 2 | 0 | 3 | 3 | 1 | Evolutionary Algorithm (C) |
|  |  |  | 521 | 9 | 0 | 1 | 14 | 33 | 24 | 18 | 35 | 32 | 35 | 54 | 13 | 3 | 9 | 1 | 1 | 2 | 18 | 0 | 0 | 3 | 4 | Genetic Algorithm (D) |
|  |  |  |  | 112 | 0 | 0 | 3 | 11 | 3 | 0 | 4 | 5 | 3 | 8 | 1 | 2 | 2 | 1 | 0 | 2 | 6 | 0 | 1 | 3 | 1 | Neural Network (E) |
|  |  |  |  |  | 101 | 0 | 0 | 1 | 0 | 0 | 1 | 0 | 0 | 0 | 0 | 6 | 4 | 6 | 1 | 7 | 1 | 0 | 1 | 0 | 2 | Semidefinite Programming (F) |
|  |  |  |  |  |  | 136 | 4 | 5 | 4 | 0 | 5 | 2 | 1 | 1 | 0 | 0 | 1 | 1 | 1 | 2 | 3 | 0 | 3 | 3 | 8 | Algorithm (G) |
|  |  |  |  |  |  |  | 219 | 10 | 7 | 1 | 23 | 21 | 13 | 10 | 3 | 6 | 0 | 2 | 6 | 2 | 4 | 6 | 6 | 1 | 6 | Scheduling (H) |
|  |  |  |  |  |  |  |  | 272 | 7 | 1 | 30 | 19 | 8 | 17 | 2 | 6 | 6 | 2 | 3 | 3 | 5 | 0 | 1 | 2 | 4 | Optimization (I) |
|  |  |  |  |  |  |  |  |  | 234 | 8 | 31 | 34 | 23 | 13 | 6 | 3 | 8 | 3 | 2 | 2 | 16 | 1 | 2 | 4 | 0 | Local Search (J) |
|  |  |  |  |  |  |  |  |  |  | 140 | 3 | 12 | 2 | 2 | 4 | 4 | 2 | 0 | 1 | 1 | 5 | 1 | 0 | 2 | 2 | Multi-Objective Optimization (K) |
|  |  |  |  |  |  |  |  |  |  |  | 419 | 28 | 35 | 17 | 3 | 24 | 8 | 4 | 9 | 2 | 12 | 1 | 1 | 7 | 8 | Heuristic (L) |
|  |  |  |  |  |  |  |  |  |  |  |  | 435 | 48 | 22 | 8 | 11 | 15 | 3 | 2 | 1 | 18 | 0 | 1 | 4 | 3 | Metaheuristic (M) |
|  |  |  |  |  |  |  |  |  |  |  |  |  | 257 | 34 | 3 | 2 | 13 | 1 | 2 | 0 | 3 | 1 | 0 | 1 | 3 | Tabu Search (N) |
|  |  |  |  |  |  |  |  |  |  |  |  |  |  | 310 | 9 | 3 | 8 | 3 | 0 | 2 | 13 | 0 | 3 | 1 | 1 | Simulated Annealing (O) |
|  |  |  |  |  |  |  |  |  |  |  |  |  |  |  | 84 | 0 | 1 | 0 | 0 | 0 | 4 | 0 | 0 | 2 | 1 | Particle Swarm Optimization (P) |
|  |  |  |  |  |  |  |  |  |  |  |  |  |  |  |  | 225 | 1 | 5 | 8 | 9 | 7 | 3 | 5 | 7 | 6 | Integer Programming (Q) |
|  |  |  |  |  |  |  |  |  |  |  |  |  |  |  |  |  | 101 | 0 | 4 | 0 | 12 | 1 | 3 | 0 | 0 | Quadratic Assignment Problem (R) |
|  |  |  |  |  |  |  |  |  |  |  |  |  |  |  |  |  |  | 84 | 5 | 3 | 1 | 1 | 0 | 1 | 3 | Lagrangian Relaxation (S) |
|  |  |  |  |  |  |  |  |  |  |  |  |  |  |  |  |  |  |  | 136 | 2 | 6 | 1 | 1 | 5 | 7 | Branch-and-Bound (T) |
|  |  |  |  |  |  |  |  |  |  |  |  |  |  |  |  |  |  |  |  | 84 | 2 | 1 | 2 | 0 | 0 | Linear Programming (U) |
|  |  |  |  |  |  |  |  |  |  |  |  |  |  |  |  |  |  |  |  |  | 225 | 1 | 7 | 3 | 4 | Traveling Salesman Problem (V) |
|  |  |  |  |  |  |  |  |  |  |  |  |  |  |  |  |  |  |  |  |  |  | 83 | 7 | 4 | 6 | Robust Optimization (W) |
|  |  |  |  |  |  |  |  |  |  |  |  |  |  |  |  |  |  |  |  |  |  |  | 108 | 0 | 3 | Computational Complexity (X) |
|  |  |  |  |  |  |  |  |  |  |  |  |  |  |  |  |  |  |  |  |  |  |  |  | 84 | 9 | Knapsack Problem (Y) |
|  |  |  |  |  |  |  |  |  |  |  |  |  |  |  |  |  |  |  |  |  |  |  |  |  | 128 | Dynamic Programming (Z) |

*Figure 6: Correlation matrix of the most relevant keywords in the research field of combinatorial optimisation. The numbers indicate how often the keywords appear together in publications. The darker the fields in the matrix are coloured green, the more often these keywords appear together.*

### 3.4.3. Application areas for combinatorial optimisation

In addition to the keyword analysis, an analysis with single words has also been conducted with the MATLAB algorithm. In this case, the keywords are divided into individual words, i.e. *combinatorial optimization*, for example, into *combinatorial* and *optimization*. This is done in order to identify important application areas for CO methods. Table 3 in the Online Appendix lists the single keywords that appear in at least 1% of publications on CO. In this table, the words that are identified as specific real-world application areas of CO (and not specific problems or methods as in Figure 5) are written in bold and the most relevant keywords and sources for the articles of these application areas are shown. The articles of a potential application area were checked by a manual inspection. For example, *scheduling*, *traveling* or *vehicle* could represent application areas, but these keywords refer almost exclusively to the CO problems scheduling problem, TSP or the vehicle routing problem, respectively.

A total of twelve application areas were identified (cf. keywords written in bold in Table 3 in the Online Appendix). The following four areas are covered most frequently (in 2% of all 8,393 articles each): *production*, *data*, *power* and *management*. The research field of *production*, which is about *production planning* (e.g. Shishvan and Sattarvand (2015)) or *assembly* (e.g. Becker and Scholl (2009)), for example, is also strongly linked to the application field *manufacturing* (e.g. Amen (2006)). Companies nowadays operate in global production networks (Lanza et al., 2019), which is the consequence of intense offshoring, outsourcing, global procurement, and expansion into new international markets. Therefore, the global production network of a typical multinational manufacturing company today includes plants spread across the globe, each facing increasing pressure to coordinate its operations with one another and with the rest of the supply chain (Ferdows et al., 2016). Research in the field of CO on the application area of production also has a comparatively high mean publication year (2011.1), which demonstrates the increasing need for novel problem solutions in this area.

The research field of *data* is linked to *information* and these two application areas mainly focus on *data mining* (e.g. Brandner et al. (2013)) or *information theory* (e.g. Braun et al. (2017)), respectively. New approaches are needed in these areas, as the flood of data in recent decades has exceeded the ability to process, analyse, store and understand the data sets. A good example are web pages whose number has increased from 1 million to 1 trillion between 1998 and 2008 alone (Fan and Bifet, 2013). The associated increasing use of data mining technologies also has a direct influence on the application field of information: for example, the increasingly emerging approaches of privacy-preserving data mining aim to protect sensitive information of individuals (Xu et al., 2014).

Big data is also related to the energy sector, which is increasingly using smart meters (sensor and measurement devices in smart grids) to collect data on real-time electricity consumption in order to better forecast and shift electrical loads (Wen et al., 2018). Energy sector related issues, which are often NP-hard (Goderbauer et al. 2019), seem to be the most frequently represented subjects in publications on CO, with the keywords *power* and *energy*. The topic of *power* is mainly about the above mentioned *smart grid* (e.g. Meskina et al. (2018)) applications or *optimal power flow* (e.g. Abido (2002)) calculations. Relevant topics in the field of *energy* include *energy efficiency* (e.g. Alharbi et al. (2019)) and *energy consumption* (e.g. Weinand et al. (2019)). In the course of the energy system transition to reduce anthropogenic greenhouse gas emissions, these issues are becoming increasingly important. This is also reflected by the mean publication year of the application area *energy* (2011.7), which is the most actual among the 12 identified application fields (cf. Table 3 in the Online Appendix). Due to the developments in industrial production already discussed above, the greenhouse gas emissions increase further, therefore energy research is also increasingly connected with the application field of production. Due to improved technologies and companies, which are increasingly trying to make their production planning energy efficient, the emissions do not increase linear. Some reasons for this from a company's perspective are policy/legislation, scarcity of resources, rising energy prices and an increasing environmental awareness (Biel and Glock, 2016).

The last two major fields of application which are covered in more than 100 CO articles are *management* with many *supply chain management* studies (e.g. Mohammadi Bidhandi et al. (2009)) and *decision* with mainly studies about *decision support systems* (e.g. Haastrup et al. (1998)). Supply chain management studies are obviously strongly tied to production and decision support is actually needed in each of the application areas.

The analysis demonstrates the importance of CO methods, which are applied in many different fields and are used to solve many of the current global problems. In addition, it is noticeable that in many of these application areas, genetic algorithms are mainly used to solve the underlying problems. This metaheuristic already turned out to be the most relevant in the area of CO (cf. Section 3.4.2). Furthermore, this single word analysis also confirms the presumptions that *robust optimisation* is currently the most prominent topic on CO, since the two strongly related terms *robust* and *uncertainty* show the highest mean publication years 2013.8 and 2013.3 respectively (cf. Table 3 in the Online Appendix). The consideration of uncertainties also becomes increasingly relevant for the real-world application areas discussed above like production planning (e.g. inhomogenety of products (Mundi et al., 2019)) or energy system analysis (e.g. stochastic nature of renewables and unknown future global energy and economy outlook (Mavromatidis et al., 2018)).

## 4. Discussion

In this section, the results are reflected upon regarding the challenges and prospects of the research field, separated for (meta)heuristics (cf. Section 4.1) and exact algorithms (cf. Section 4.2). Furthermore, the limitations of the study are discussed in Section 4.3.

### 4.1. Challenges and prospects of (meta)heuristics

One overarching conclusion from the data presented in this paper is that a large majority of papers in the field of CO are still of the "problem-algorithm-results" type. In these papers, an algorithm for a specific CO problem is developed and tested, and in most cases demonstrated to perform well by comparing it to other algorithms for the same problem. Several authors (Barr et al., 1995; Hooker, 1995; Kendall et al., 2016) have warned that this type of research, especially the competitive testing aspect, yields very little scientific *knowledge* beyond the anecdotal. It wastes enormous amounts of research time on "development" activities (polishing code, compiler tuning, …) necessary to achieve top-notch performance. Papers in which attempts are made to draw some generalizable conclusions on heuristics and metaheuristics exist (e.g. Santini et al. (2018) or Watson et al. (2003)), but still represent a marginal phenomenon.

The practice of deciding which papers to publish based on competitive testing (which has been called the "horse race") has also resulted in a considerable *publication bias*. Mainly positive results (algorithm X works well for problem Y) appear in the literature and negative results, that demonstrate that some type of heuristic does *not* work for some problem (category) are few and far between (although some exist, e.g. Sörensen and Schittekat (2013)). Combined with the fact that there are very few generally accepted protocols for testing and reporting on algorithmic results, most academic journals do not even require simple statistical tests to demonstrate that a "better performance" is significant in the statistical sense. This makes one wonder how well many of the results published in the literature would stand the scrutiny of independent replication and testing (see, e.g. Sörensen et al. (2019) for an example where an independent replication was not able to confirm the authors' original performance claims).

Another observation is that there seems to be a widening divide between the communities on metaphor-based metaheuristics (also called "nature-inspired" metaheuristics, even though the inspiration for this category of metaheuristics now comes from sources that have little to do with nature), and the more traditional metaheuristics that are not based on some metaphor. In the field of metaphor-based metaheuristics, the decision on whether an algorithmic idea is valuable or not seems to hinge on the novelty of the metaphor that inspired it, with ever more outlandish metaphors being proposed (interior decoration (Gandomi, 2014), the FIFA world cup (Razmjooy et al., 2016), "intelligent" water drops (Hosseini, 2009), and – in a demonstration of spectacular opportunism – the spread of covid-19 (Martínez-Álvarez et al., 2020))³. Without implying that the field of traditional metaheuristics does not have its issues, one can only conclude that the scientific standards in the subfield of metaphor-based metaheuristics are particularly low. As an illustration, it is remarkable that one of the top-cited papers in this field is the paper introducing the Harmony Search algorithm (Geem et al., 2001) (cf. Section 3.3). This metaheuristic is supposedly based on musicians playing music together (a solution in Harmony Search lingo is called a "melody", for example), even though this algorithm has been unequivocally demonstrated (Weyland, 2010, 2015) to be a special case of Evolution Strategies, a metaheuristic that predates it by 30 years. The community project "Metaheuristics in the Large" (MitL) has recently formulated a framework that enables combinatorial assembly and comparison of metaheuristics, and thereby also addresses issues of reproducibility and scalability (Swan et al. 2020).

Another trend, that is perhaps difficult to glean from the bibliometric results in this paper is that there is an increasing focus on "rich" problems (i.e., problems with complex formulations that involve many specific constraints and objectives), often based on real-life applications. Without doubt, heuristic and metaheuristic ideas have penetrated into the mindsets of practitioners that develop software for real-life optimization, and increasingly, (meta)heuristic research is finding its way to practical application. As an illustration, (software) companies like PTV and ORTEC increasingly participate in conferences and write out challenges for researchers to solve their real-life problems (Kheiri et al., 2019).

Finally, reserachers in the field of (meta)heuristics are increasingly reaching out to related fields, incorporating ideas and techniques to develop better optimization algorithms. The combination of heuristics with exact methods (often called "matheuristics") has developed into a field of its own. Combinations with constraint programming and machine learning are also increasingly being found. For a concise review of such combinations, please refer to (Talbi, 2016).

### 4.2. Challenges and prospects of exact algorithms

The main keywords concerning publications on exact CO algorithms, according to the data reported in this work, are *branch-and-bound* and *integer programming*. Also from the data, the number of publications related to exact algorithms is considerably less than those related to approximation algorithms. One reason for this is that many hard combinatorial problems of practical interest are NP-hard and only approximation methods are able to provide *good* solutions. However, exact algorithms for NP-hard problems which are able to solve a number of real networks instances with millions of nodes to proven optimality, have recently been described (e.g., San Segundo et al. (2016) or Walteros and Buchanan (2020)).

Another explanation for the relatively small number of publications concerning exact CO algorithms might lie in the already mentioned "problem-algorithm-results" structure of the papers. Typically, this line of work focuses on the study of the specific structure of problem instances and develops specialized new algorithms, which are then compared extensively with current state-of-the-art approaches. Only algorithms which show a significant improvement over state-of-the-art get to be published in the top CO journals. In addition, it has been argued that exact algorithms should also be *certifying*, i.e., provide an easily verifiable proof that the solution is correct (Gocht et al., 2020).

---
³ The interested reader is referred to the "EC Bestiary", a satirical compilation of all metaphor-inspired metaheuristics published in the literature (Campelo and Aranha, 2019).

Notwithstanding, this "problem-algorithm-results" stream of research has shown some exciting improvements for some fundamental NP-hard / NP-complete problems, such as the maximum independent set problem and the boolean satisfiability (SAT) problem. Specifically for the latter and during the past 20 years, the progress on the algorithmic methods has been reported to have at least the same impact as the advances in hardware (Fichte et al., 2020). In addition, research on specific NP-complete problems has also led to the study of effective transformations between problems. Among the many examples of recent successful transformations, SAT modules of state-of-the-art constraint programming solvers can be pointed out (e.g. Zhou et al. (2015)) as well as the SAT-based bounding functions used by state-of-the-art maximum clique solvers (Li et al., 2018). Complementary to the development of new exact algorithmic techniques are the numerous CO challenges that are organized periodically, such as XCSP3, PACE, DIMACS and SAT competitions, where algorithms are compared against benchmarks of practical interest and the state-of-the-art is settled.

From the data available, it is also worth mentioning the study of methods which aim at speeding up the convergence of exact CO algorithms by reducing the problem instance exploiting structural properties. These techniques can be applied just once, in a pre-processing phase known as *kernelization*. Alternatively, they can be applied at every node of the branching tree, denoted the *branch-and-reduce* paradigm. Both techniques have recently contributed to solve to proven optimality some hard combinatorial large scale problem instances (Akiba and Iwata, 2016; Hespe et al., 2020). Future research in exact CO algorithms capable of solving large scale real life problems will definitely continue over time. Exciting developments are to be expected from new algorithmic branch-and-bound and branch-and-reduce techniques, but one may also look forward to new algorithmic frameworks, of which the recent *branch-cut-and-prize* framework for vehicle routing and other related problems are good examples (Pessoa et al., 2020).

Since these challenges and prospects refer mostly to the results of Section 3, other emerging issues are likely to be neglected. One example would be machine learning based approaches, which show promising results in different applications by making decisions that were otherwise made by handcrafted expert knowledge based heuristics in a more principled and optimized way (Bengio et al., 2020). For example, in the original AlphaGo paper (Silver et al., 2016), a machine learning algorithm is first trained based on expert knowledge and refined in a further step by using a reward signal from games of self-play (reinforcement learning). In general, machine learning is used in both exact and heuristic frameworks (Bengio et al., 2020).

### 4.3. Limitations of this study

As with all review studies, the evaluation of the topic in this bibliometric analysis depends on the type of search query. Most probably, a large number of studies cover CO problems, but do not use combinatorial optimization as a keyword, nor in title, abstract or article. If these publications are also not assigned to CO via KeywordPlus in Web of Science, they are not included in the present analysis. Thus, a 100% comprehensive picture of CO cannot be given. On the other hand, the objective is to analyse studies that explicitly deal with CO as a methodology and the analysed corpus of literature is assumed to yield a representative sample of the CO research field.

Furthermore, some methods use parameters that are set in a more or less arbitrary way, and different parameters may yield different results. Examples are the number of years in the trend analysis, or the maximum distance set for the Levenshtein distance. It is not possible to recognize all relevant keywords during the grouping process using distances like the Levenshtein distance, because this always involves some interpretation. For example, Figure 5 shows the keyword traveling salesman problem, which is used 225 times. The higher occurrence (cf. Table 3 in the Online Appendix) of the words traveling (384) and salesman (349), however, suggests that the traveling salesman problem is covered in more than 225 articles. However, not all articles can be grouped by an algorithm if significantly different keywords are used. For example, unlike simulated annealing, which is always known under that name, large neighborhood search is also known as destroy-and-repair and ruin-and-recreate, but the automatic analysis does not (and cannot) group these concepts. Therefore, the keyword analysis in Section 3.4 is only to be understood as an indication of trends.

## 5.  Summary and conclusions

More and more real-world problems are becoming highly complex and have to be solved with combinatorial optimisation techniques, which have always been of great interest to the scientific community. Consequently, the number of publications on combinatorial optimisation has increased exponentially between 1990 and 2019, amounting to 8,393 at the time of this analysis. This requires a study of the corpus of literature to show the status quo and trends in research on combinatorial optimisation. The present study therefore uses a bibliometric analysis, supported by the literature database Web of Science, the R-tool bibliometrix as well as a novel algorithm developed for keyword analyses.

Among the 85 contributing countries, the USA is the most important contributor with 1,918 articles and the highest h-index (104), followed by the China (1,197 articles), with the highest annual number of publications since 2014, and France (676 articles). In general, the share of collaborative publications on combinatorial optimisation studies is rather low, with the most collaborations between USA and China (146). The most productive organisations are the *French National Centre for Scientific Research* (235 articles), the *University Of California System* with 162 articles as well as the highest h-index (37) and the *Chinese Academy of Sciences* (118 articles). Due to the low proportion of cross-country collaborations, organisations from the same country tend to collaborate, with the *University of Montreal* and *Polytechnique Montreal* recording the highest number of collaborations among institutes (42). Core sources on combinatorial optimisation are *European Journal of Operational Research*, *Lecture Notes in Computer Science* and *Computers & Operations Research*, which published around 15% of the 8,393 articles.

The analysis of the most relevant publications and author keywords shows that the majority of studies focuses on the development, extension and application of metaheuristics. While in most cases genetic algorithms predominate (6% of the 8,393 publications), metaheuristics are mostly tested or applied to the Traveling Salesman problem. Among the most globally cited publications are several articles on ant colony optimisation. It appears that in the past the development of metaheuristics required only a new type of metaphor to justify its development. In this context, there seems to be a widening divide between the communities on metaphor-based metaheuristics (also called "nature-inspired" metaheuristics) and the more traditional metaheuristics that are not based on some metaphor. However, this trend of algorithm orientation in metaheuristics seems to have partly changed to a more problem-oriented approach in recent years. In this context, the analysis of the keywords showed that combinatorial optimisation problems are particularly relevant in real-world application areas in the energy sector, production sector and data management. This is due to the need to solve complex problems related to global production networks, the reduction of greenhouse gas emissions in the course of energy system transition and the increasing amount of big data and its processing. The most current topics in the research area of combinatorial optimisation are uncertainties and the associated increasingly relevant methodology of robust optimisation, which is also becoming more and more important in the aforementioned application areas. A discussion of the challenges and prospects of the field further reveals that most articles are still of the "problem-algorithm-results"-type, which requires large amounts of research time on "development" activities necessary to achieve top-notch performance.

The present bibliometric analysis demonstrates global research trends in combinatorial optimisation. This study can therefore support the scientific community as well as policy makers in identifying the relevant issues regarding the expanding and transforming combinatorial optimisation research area and its real-world applications.

**Acknowledgements**

This work has been partially funded by the Spanish Ministry of Science, Innovation and Universities through the project COGDRIVE (DPI2017-86915-C3-3-R).